\documentclass[aps,pre,preprint,superscriptaddress,showkeys]{revtex4}
\usepackage{graphicx}
\usepackage{times}
\usepackage{amsmath}
\usepackage{amssymb}
\begin{document}
\def\T{\Theta}
\def\D{\Delta}
\def\d{\delta}
\def\r{\rho}
\def\p{\pi}
\def\a{\alpha}
\def\g{\gamma}
\def\ra{\rightarrow}
\def\s{\sigma}
\def\b{\beta}
\def\e{\epsilon}
\def\G{\Gamma}
\def\om{\omega}
\def\pe{$1/r^\a$ }
\def\l{\lambda}
\def\f{\phi}
\def\w{\psi}
\def\m{\mu}
\def\t{\tau}
\def\c{\chi}

\title{On the evolution of decoys in plant immune systems}
\author{Iaroslav Ispolatov \& Michael Doebeli \\
\vspace{-2mm}\normalsize Department of Zoology and Department of Mathematics, \\
\vspace{-2mm}\normalsize University of British Columbia, Vancouver B.C., Canada
V6T 1Z4\\
}

\begin{abstract} The Guard-Guardee model for plant immunity describes how resistance proteins (guards) in host cells monitor  host target proteins (guardees) that are manipulated by pathogen effector proteins. A recently suggested extension of this model includes decoys, which are duplicated copies of guardee proteins, and which have the sole function to attract the effector and, when modified by the effector, trigger the plant immune response. Here we present a proof-of-principle model for the functioning of decoys in plant immunity, quantitatively developing this experimentally-derived concept.  Our model links the basic cellular chemistry to the outcomes of pathogen infection  and resulting fitness costs for the host. In particular, the model allows identification of conditions under which it is optimal for decoys to act as triggers for the plant immune response, and of conditions under which it is optimal for decoys to act as sinks that bind the pathogen effectors but do not trigger an immune response.
\end{abstract}


\maketitle
\section  {Introduction} 

The Guard-Guardee model of pathogen recognition has become an established paradigm in plant immunity \cite{bomblies_etal2007, jones_dangl2006, ispolatov_doebeli09, hoorm_kamoun_09}. According to this model, an effector molecule, injected by the pathogen into a host cell to facilitate infection, targets some molecule that is important for cellular life. The state of this guardee molecule is monitored by a specific immune resistance (R) protein, also called a guard.  When the state of 
the guardee molecule is modified by the effector, the R protein detects this change and activates an immune reaction that usually kills the cell, thus preventing pathogen proliferation and warning other cells of the threat of infection. However, based on recently acquired empirical data, it has been conjectured, that the plant innate immune system employs even more sophisticated mechanisms to fight off pathogen attacks.

In particular, \cite {hoorm_kamoun_09} have suggested that a duplication of some of the effector target genes and subsequent  independent evolution of the duplicated genes can create decoys of the effector targets. Such a decoy 
is free from the non-immune cellular functions of the original target proteins and can evolve to become an effective bait for the pathogen effector, as well as a better trigger for its corresponding guard R protein. Meanwhile, in the presence of decoys the target protein might have fewer constraints to mutate away from the recognizability by the effector, which might be beneficial to the individuals without functioning R genes.
Four examples of possible effector perception mechanisms, summarized in \cite{hoorm_kamoun_09}, support 
the decoy model. The diversity of these examples, which include the Pto protein as a candidate for the decoy in tomato, pBS3 in pepper, RCR3 in tomato as a bait for fungus effector, and the RPS2 protein in Arabidopsis, indicate that such decoys may often evolve independently and may be active in a wide variety of plant-pathogen interactions.  

It is interesting that the notion of decoy is also being used in the different, but related context of ``decoy receptors'', which are  inactive duplicates of normal receptors for general signalling pathways. Such decoys have high and selective binding affinity to specific ligands but lack the signal transduction domains of normal receptors.  Decoy receptors act as sinks for ligands and  prevent them from binding to their active targets (see, for example, \cite{liberto08,ashkenazi02}  or work as parts of multiprotein receptor complexes, \cite{leblanc03}. Importantly, it has  been reported that certain cancer cells evolve to escape immune-cytotoxic attack by over-expressing a decoy receptor that sequesters the respective apoptosis ligand \cite{pitti98}.
For immune reactions in plants, it therefore seems natural to assume that decoys can function in two different ways: first, as being baits for triggering the guard-mediated immune response and, second, in case of sufficient abundance and not too strong recognition by guards, by simply absorbing the pathogen effectors and preventing them from attacking their normal targets, thus preventing both infection and the immune response.  In the following we will often refer to these two mechanisms and decoy roles as a ``trigger'' and ``sink''.

Our goal here is to quantify the presently only qualitative description of both trigger and sink mechanisms of decoy functioning in host immune response, and to understand the conditions that might favour one or the other mechanism evolutionarily. The advantages and disadvantages of each of the mechanisms are summarized in  the form of the overall fitness decrease of the organism caused by the pathogen infection. 
We develop a simple  model which links some basic cellular chemistry to probabilities of possible outcomes of parasite attack and their fitness costs for a host organism. 
In our model we consider reversible binding interaction between four major types of molecules: pathogen effector, its target in a plant cell (guardee), the decoy of this target, and the guard R protein. The effector can bind both to the target and the decoy, and the guard protein can interact with the effector-target and effector-decoy complexes and thus trigger the immune response. Fitness costs are associated with the maintenance of decoy proteins, with cell death caused by the activated immune response, and with undetected and therefore successful infection, which also causes cell death, but enables pathogen multiplication and subsequent attack on a number of adjacent cells. We find that decoys are typically favoured evolutionarily, but depending on the parameters of the model, the decoy can indeed exhibit either the role of a sink for effector molecules, or the role of an effective trigger of the immune response. In particular, when the cost of maintenance of the decoy is low, higher fitness is achieved when decoys act as a sink: the decoy protein tends to be more abundant and simply sequester the effector without being recognized by the guard, thus avoiding cell death. However, when the cost of the maintenance of the decoy is elevated, higher fitness is achieved with the fewer decoy molecules being triggers of the immune response, and with a high recognizability of the effector-decoy complex by the guard protein.   

The structure of the paper is the following: In the next section we describe our model, starting with the basic cellular chemistry, linking the chemical concentrations to the probabilities of outcomes for the infected plant, evaluating the fitness costs of each outcome, and determining parameters and scenarios that  maximize the fitness. The discussion and conclusion sections completes the paper.

\section{ The model}
\subsection{Binding equilibrium}
We consider biochemical reactions in a system initially consisting of 4 types of monomers (proteins): pathogen {\it effector}, denoted by $E$,  {\it effector target} (guardee) in a host cell, denoted by $T$,  {\it decoy of the effector target} denoted by $D$, and a {\it guard R protein} denoted by $R$, which monitors the state of the effector target and its decoy. These monomers can reversibly react in the following ways:
The effector can bind both to its target to form a complex $ET$ and to the decoy to form a complex $ED$; the guard protein can recognize modified effector targets  and modified decoys by binding to $ET$ and $ED$ complexes, thus 
forming $RET$ and $RED$ three-molecular aggregates. It is assumed that guard proteins do not bind to non-modified targets or decoys, and hence that the formation of complexes $RE$ and $RD$ is impossible.
We assume that the chemical equilibrium between these binding states establishes itself fast and that only the steady state concentration of complexes are biologically relevant. We also assume that the concentrations of all reactants are uniform and therefore disregard any spatial dependence.
The equilibrium concentrations of the constituent proteins and their complexes are uniquely defined by the law of Mass Action, (\ref{lma}),
\begin{eqnarray}
\label{lma}
 [E] [T] = K_{ET} [ET]\\
\nonumber
 [E] [D] = K_{ED} [ED]\\
\nonumber
[R] [ET] = K_{RET} [RET]\\
\nonumber
[R] [ED] = K_{RED} [RED].
\end{eqnarray}
Here $[A]$ denotes the concentration of substance $A$, and $K_{AB}$ is a dissociation constant of a complex $AB$.
A dissociation constant has the dimensionality of a concentration, and a smaller dissociation constant means stronger binding. 
In addition, there are conservation laws which state that the total amount of each of the four proteins, i.e. the amount of free and bound protein, remains constant and is equal to $[E]_0$, $[T]_0$, $[D]_0$, and $[R]_0$, respectively. We assume that the values of the total concentrations change very slowly on the timescale of the development and detection of a pathogen attack and therefore approximate them by constant values.
\begin{eqnarray}
\label{sum}
 [E]_0 = [E] + [ET] + [ED] + [RET] + [RED]\\
\nonumber
 [T]_0 =[T] + [ET] + [RET] \\
\nonumber
[D]_0 = [D] + [ED] + [RED] \\
\nonumber
[R]_0= [R] + [RET] + [RED].
\end{eqnarray}
The system of 8 equations resulting from (\ref{lma},\ref{sum}) in principle allows to determine all 8 equilibrium concentrations, but the equations are nonlinear and can be solved only numerically. A simple way to numerically find the unique equilibrium concentrations \cite{shear68} is to integrate the actual kinetic equations that lead to the establishment of the equilibrium over time, keeping the ratio of the kinetic coefficients for association and dissociation equal to the corresponding dissociation constants \cite{maslov07}.
For example, for a complex $AB$ formed as a result of the reaction
\begin{equation}
A + B  \rightleftarrows AB,
\end{equation}
with a dissociation constant $K_{AB}$, the corresponding system of kinetic equations is
\begin{eqnarray}
\frac{d[A]} {dt} = - [A][B] + K_{AB} [AB] \nonumber \\
\frac{d[B]}{dt} =  -[A][B] + K_{AB} [AB] \\ \label{kinetics}
\frac {d[AB]}{dt} = [A][B] - K_{AB}[AB]. \nonumber 
\end{eqnarray}
For each protein, the corresponding equations are numerically evolved in time until the steady state concentrations are  reached.

\subsection{Probability of effector and guard success and concentration of complexes}
The steady state concentrations of protein complexes are directly related to  probabilities of success or failure of the pathogen effector to induce changes in its target or decoy, and to the probability that a guard protein  recognizes these changes. We assume that the extent to which the effector modifies the cellular environment to the benefit of the parasite depends on the fraction of target molecules that it binds to. 
 Similarly, the efficiency of recognition of the effector attack by the guard protein depends on the fraction of guard protein molecules bound to the effector-target complex. While generally the dependencies between the probabilities of such effects and concentrations ratios can have a complex functional form, in the following we assume a linear dependence between the probability of an outcome and the fraction of corresponding monomers bound into a complex. 
These fractions of concentrations are naturally bound between zero and one and thus can be linked to probabilities without any need for further renormalization; also, such a linear dependence is the simplest form of a monotonic map. 
Consequently, the fraction of target molecules that is bound to the effector, whether such binding is recognized by the guard or not,
$P_T=([ET] + [RET])/[T]_0$, quantifies the probability of manipulation of  the cellular chemistry by the effector. The fraction of the guard protein molecules bound to the effector-target complex, $Q_T=[RET]/[R]_0$, reflects the efficiency of recognition by the guard protein of such manipulations. Likewise, the fraction of the guard protein concentration bound to the effector-decoy complex,  $Q_D=[RED]/[R]_0$, describes the efficiency of recognition by the guard protein of the changes induced by the effector in the decoy.  The binding between effector and decoy by itself does not cause any physiological consequences, hence we do not introduce any special notation for the probability of such events. Note that our description takes into account the possibility of an "overreaction" by the plant immune system, that is, a possible immune response induced by the guard protein when $Q_T$ or $Q_D$ is large (close to one), while $P_T \ll 1$, i.e. when the  the actual infection of the cell is rather improbable.

\subsection{Probabilities of outcomes} 
After the pathogen has injected the effector into a cell, three distinct scenarios can develop: 
\begin{itemize}
\item Nothing happens to the cell
\item The immune system kills the cell and the infection does not spread
\item The parasite successfully replicates in the cell and subsequently infects other cells
\end{itemize}

In terms of the quantities $P_T$, $Q_T$, and $Q_D$ the probability that nothing happens to a cell in the presence of the effector is
\begin{equation}
\label{nothing}
\Pi_N = (1-P_T)(1-Q_T)(1-Q_D).
\end{equation}
This expression means that the effector neither manipulates its target sufficiently to cause the infection, nor does it change the target and decoy sufficiently for either change to be recognized by the guard protein. 
Similarly, the probability of infection is given by
\begin{equation}
\label{infection}
\Pi_I = P_T(1-Q_T)(1-Q_D).
\end{equation}
Here the effector does manipulate the target sufficiently to cause the infection, yet the changes in target and decoy are not recognized by the guard. 
Finally the probability of activation of  the immune response and killing the cell, and preventing the spread of infection is 
\begin{equation}
\label{IR}
\Pi_{IR} = Q_T + Q_D - Q_T Q_D.
\end{equation}
The first two terms on the right hand side describe the probabilities of immune response induced either by the effector target or the decoy, and the third term eliminates double counting in the event when both the target and decoy are inducing the immune response.  
These three scenarios form a complete set of events, so 
$$
\Pi_N + \Pi_I + \Pi_{IR} =1.
$$

\subsection {Fitness costs}
Now we evaluate fitness costs of these three possible scenarios.  First we assume that the cost of maintenance of  the decoy protein reduces the fitness of an organism by the amount proportional to the concentration of this protein, 
\begin{equation}
\label{cm}
\sigma _M= - C_M \frac{[D]_0}{[T]_0} 
\end{equation}
Minus signifies the reduction of fitness, and $[T]_0$ in the denominator sets the concentration scale.
An induced immune response implies the death of the attacked plant cell, so the fitness cost of the immune response is the cost of replacement of the cell, which we denote by $\sigma _{IR}$.

Finally, the fitness cost of the infection can be expressed through $\sigma_{IR}$ and  a new parameter
$\mu$, which denotes the number of new cells that become subjected to parasite attack as a consequence of successful parasite multiplication in the infected cell. 
\begin{equation}
\label{ci}
\sigma _{I}= \sigma_{IR} + \mu( \Pi_I \sigma_I + \Pi_{IR}\sigma_{IR}).
\end{equation}
The first term on the right-hand side of this self-consistent equation reflects the fitness cost due to the death of the infected cell. The second term takes into account that each of $\mu$ cells subject to parasite attack due to infection of a primary cell either become infected themselves with probability $\Pi_I$ and fitness cost  $\sigma_I$, or they successfully thwart the pathogen attack with probability $\Pi_{IR}$ and at a cost $\sigma_{IR}$ due to activation of the immune response.
When $\mu \Pi_I > 1$, equation (\ref{ci}) has no positive solution for $\sigma_I$, which describes the indefinite spread of infection so that all cells in the organism die.
However, when  $\mu \Pi_I < 1$, the organism eventually overcomes the infection with the total cost to fitness
\begin{equation}
\label{ci2}
\sigma _{I}= \sigma _{IR} \frac{1 + \mu \Pi_{IR}}{1 - \mu \Pi_I}.
\end{equation}

Consequently the total fitness reduction caused by introduction of an effector into a cell is equal to the sum of fitness costs, Eqs.~(\ref{cm},\ref{ci2}), where the costs of immune response and infection need to be multiplied by the probability of the corresponding outcome:
\begin{equation}
\label{ct}
\sigma =- C_M \frac{[D]_0}{[T]_0}  -\sigma _{IR}\Pi_{IR}- \sigma _{IR}\Pi_{I} \left( \Pi_{IR} + \frac {1 + \mu \Pi_{IR}}{1 - \mu \Pi_I} \right).
\end{equation}

\section {Fitness optimization}
Once the fitness costs of all three possible outcomes are established, we look at how the organism can maximize fitness, or minimize total fitness loss. Since the decoy is assumed to be free of other, non-immune roles, its evolution should be the fastest and the least constrained.
Thus we look for the fitness maximum, varying the parameters related to the decoy which are assumed independent: the dissociation constants for binding between the decoy and the effector $K_{ED}$ and between R-protein and decoy-effector complex $K_{RED}$, and the decoy concentration $[D]_0$.
Unless otherwise specified, all concentrations and dissociation constants are assumed to be constant and equal to one, and $\mu=5$.

The following general evolutionary pattens were observed when the cost of decoy maintenance was varied:
\begin{itemize}
\item The binding between the effector and decoy always tends to be maximized, $K_{ED} \rightarrow 0$, so that for both the sink and trigger functions, the decoy absorbs as many effector molecules as possible.  This seems natural, as it is worthless to maintain any number of decoy copies if these copies are not interacting with the effector.
Thus in the following we set the effector-decoy dissociation constant to $K_{ED}=0.2$; the precise value of $K_{ED}$ does not affect our conclusion provided that it is noticeably less than all the relevant concentrations, and it seems reasonable to assume that evolution of $K_{ED}$ stops when the strong binding limit is reached. 
\item For high costs of maintenance of the decoy (roughly for $C_M \geq 0.3\sigma_{IR}$), the fitness optimum is reached for  smaller decoy concentrations and strong binding (small $K_{RED}$) between the R-protein and effector-decoy complex. Thus, when it is desirable to maintain fewer decoys due to high costs, decoys are predominantly used as effective triggers of the immune response, Fig.~1. 
\item For low costs of maintenance of the decoy (roughly for $C_M\leq 0.2 C_C$) the cell produces more decoy molecules; decoys simply absorb the effector without triggering the immune response, so that $K_{RED}$ is large, Fig.~2. Hence the cheaper and more abundant decoys become sinks for effectors, and save more cells from the death inevitably caused by the immune reaction. 
\item The ``survival'' curve, which separates the area of the organism death (marked as white area at the bottom of all Figures) from the area of survival with a finite fitness cost (grey area with contour lines) is practically invariant in all figures and independent on the decoy maintenance fitness cost. This means that for small decoy concentration the survival requires strong binding between the decoy-effector complex and the guard protein, or, in other words, a reliable immune trigger. 
\end{itemize}

\begin{figure}
\includegraphics[width=.4\textwidth]{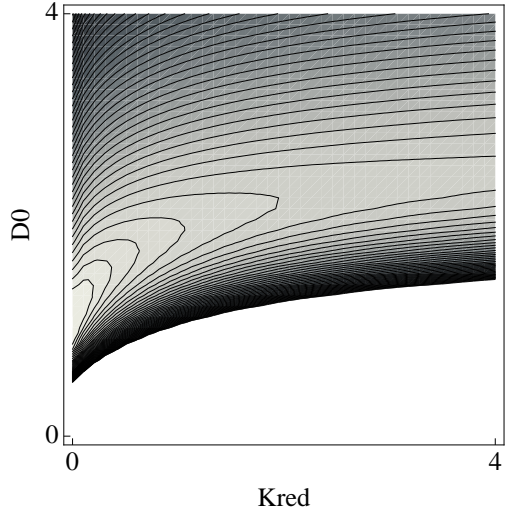}
\caption{\label{1fig} Expensive, $C_M=0.3C_C$, decoy tends to be less abundant  and develop strong binding between the effector-decoy complex and the guard protein. 
In this case  the decoy playing the role of an immune response trigger yields higher fitness. 
White area corresponds to death of the organism. 
}
\end{figure}

\begin{figure}
\includegraphics[width=.4\textwidth]{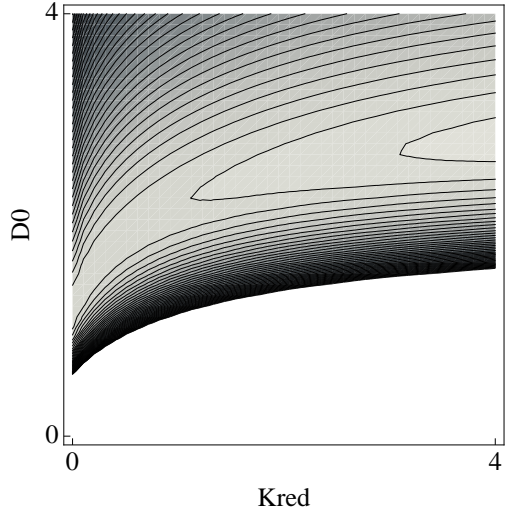}
\caption{\label{2fig} Less expensive, $C_M=0.2C_C$, decoy tends to be more abundant  and develop weaker binding between the effector-decoy complex and the guard protein. In this case the decoy's role of a sink for effector provides the higher fitness. 
}
\end{figure}

In a second set of numerical experiments we varied the binding affinity of  the guard protein to the target-effector complex, $K_{RET}$. Smaller   $K_{RET}$ correspond to a well-tuned immune responses that is sensitive to manipulation of the target by the effector, while larger  $K_{RET}$ correspond to slow and unreliable immune reaction.
We observed that:
\begin{itemize}
\item For strong recognizability ($K_{RET}=0.5$, Fig. 3), the decoy tends to play the role of a trigger, being less abundant (smaller $[D]_0$) and more recognizable for the guard protein (smaller $K_{RED}$).  In this case the probability of inducing immune response is already high without a decoy, which makes cell death highly plausible. The additional decoy improves the reliability of effector detection by the cell even more, yet the concentration of the decoy is not too high due to the finite cost of its maintenance.
\item For weak recognizability ($K_{RET}=2.$, Fig. 4), the decoy tends to be  a  more abundant (larger  $[D]_0$) sink, with the affinity for binding between the decoy-effector complex and the guard being smaller (larger $K_{RED}$).  In this case the cell has more chances to avoid death both from the immune response and infection by sequestering the effector through the decoy. The total fitness in this case is higher than in the former one. 
\end{itemize}

\begin{figure}
\includegraphics[width=.4\textwidth]{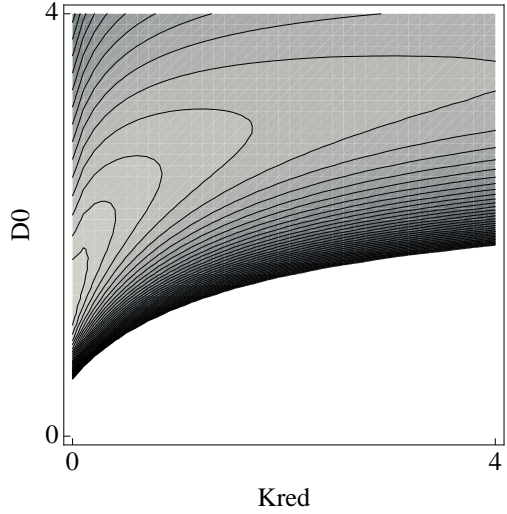}
\caption{\label{3fig}  Strong recognizability of effector action on the target by the guard, $K_{RET}=0.5$. The decoy tends to be less abundant and to  make the decoy-effector complex 
more recognizable to the guard, thus playing a role of immune response trigger.
}
\end{figure}

\begin{figure}
\includegraphics[width=.4\textwidth]{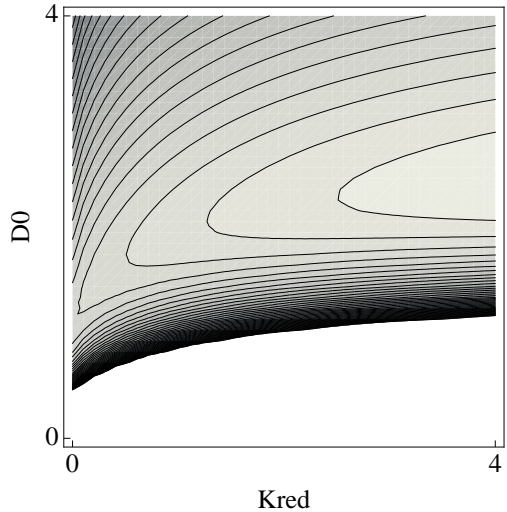}
\caption{\label{4fig}  Weak recognizability of effector action on the target by the guard, $K_{RET}=0.5$. The decoy tends to be more abundant and to make the decoy-effector complex 
less recognizable to the guard, thus playing a role of a sink for the effector.
}
\end{figure}


Finally, we considered how the severity of the pathogen attack shapes the optimal decoy strategy.
A stronger attack can manifest itself as an elevated concentration of pathogen effectors. In Fig.~5 we show that the increase in concentrations of effector to $E_0=2$ leads to  stronger recognizability (smaller $K_{RED}$)  and smaller concentration of the decoy. A similar effect occurs when the concentration of the effector is kept constant, but the effector is more ``virulent'', so that the affinity of binding between the effector and its target increases , $K_{ET}=0.5$, Fig.~6. Thus in general, an increase in severity of the pathogen attack leads to a switch of the decoy's role from being a sink to being trigger (unless the decoy already is a trigger, e.g. due to high maintenance costs).
\begin{figure}
\includegraphics[width=.4\textwidth]{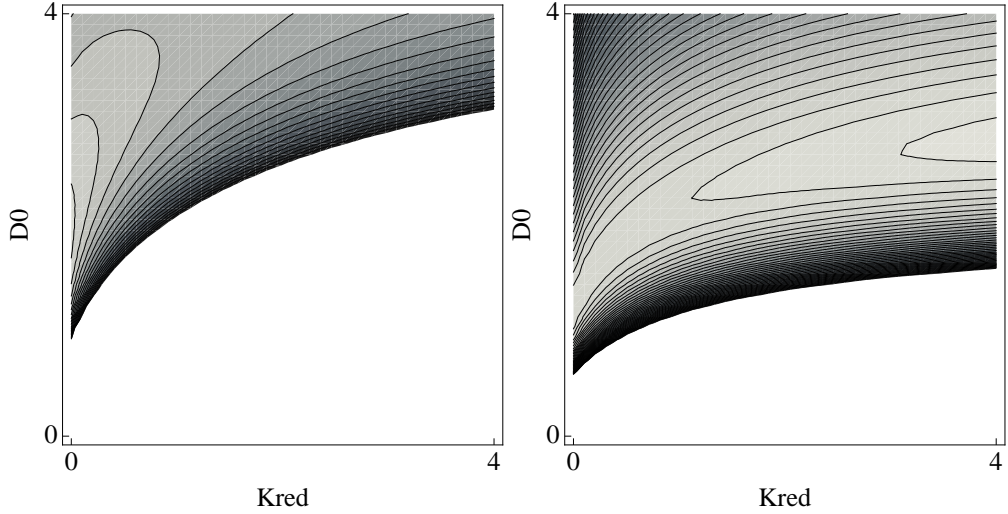}
\includegraphics[width=.4\textwidth]{f2.eps}
\caption{\label{5fig}
The higher concentration of effector, $E_0=2$, (left panel) shifts the fitness towards lower decoy concentrations $D_0$ and higher recognizability by the guard of the decoy-effector complex (smaller $K_{RED}$). Right panel shows the control system with 
$E_0=1$ (same as Fig. 1) , in both panels $C_M=0.2C_C$. 
}
\end{figure}

\begin{figure}
\includegraphics[width=.4\textwidth]{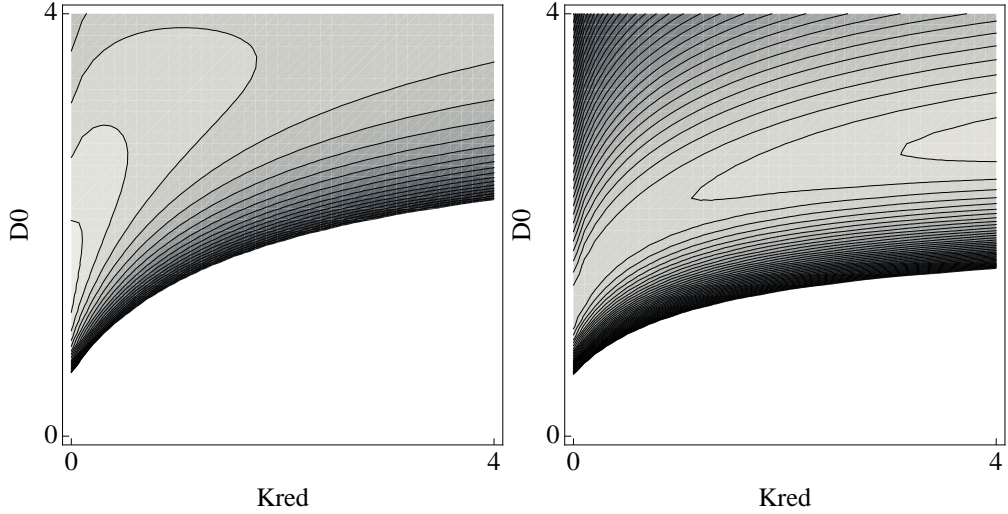}
\includegraphics[width=.4\textwidth]{f2.eps}
\caption{\label{6fig}
The stronger binding of effector to target, $K_{ET}=0.5$, (left panel) shifts the fitness maximum shifting towards lower decoy concentrations $D_0$ and higher recognizability by the guard of the decoy-effector complex (smaller $K_{RED}$). Right panel shows the control system with 
$K_{ET}=1$ (same as Fig. 1) , in both panels $C_M=0.2C_C$. 
}
\end{figure}
\section{Discussion and conclusions}

We developed a model intended to shed some quantitative light on the two basic ways in which decoy proteins are thought to function in plant immunity. 
Decoys, which are most probably duplicated copies of pathogen's effector targets in a plant cell, presumably evolved to play the sole role to efficiently mimic effector targets \cite{hoorm_kamoun_09}. However, after 
attracting the pathogen effector that invaded a plant cell, the decoys can subsequently function in two ways:
In the first role, decoys act as sinks for pathogen effector proteins, so that the pathogen attack is thwarted simply by rendering the effector proteins useless for changing the hosts cellular environment in favour of the pathogen. In the second role, the decoy protein act as triggers for the host immune response,  revealing the presence of effectors to the guard R proteins that detect  the  effector-induced changes in the  targets and decoys, thus
effectively making the host immune response more sensitive. So far, these two modes of operation were known to occur in different signaling pathways and in different biological context: The best studied examples of decoys acting as sinks 
come from cancer-related studies of decoy receptors, which
absorb cytokines and other signaling molecules active in the apoptotic signaling loop
\cite{ashkenazi02}, and are not directly related to plant immunity.
The known examples of decoy-triggers come from the rather sophisticated
extension of well-established guard-guardee plant immune system functioning mechanism \cite{hoorm_kamoun_09}. Our models suggest that the basic functioning of these two decoy modes is closely related. In particular, based on our results it can be expected that under some conditions, the decoy proteins involved in plant immunity may also play the role of simple sinks absorbing pathogen effectors. 

Our quantitative model  is based on simple yet realistic biochemical postulates and connects the reversible binding between four types of monomer molecules (pathogen's effector, effector target, decoy of effector target, and guard protein)  to the probabilities of outcomes such as cell death and immune reaction, and finally, to the total fitness costs of such outcomes. We assumed that the decoy is free to mutate to achieve the highest fitness. Depending on the cost of maintaining decoy proteins, on the efficiency of the guard proteins triggering the immune response, and on the intensity of the pathogen attack, the optimal decoy strategy may consist of being a sink or of being a trigger. When a decoy is costly to maintain, it tends to become a less abundant but a reliable immune trigger, while when the cost to express many decoy copies is low, it is more likely to be a simple sink. Also, when the immune system functions well without a decoy and the probability of detection of pathogen attack is high, the decoy tends to further improve the reliability of triggering the immune reaction rather than to simply absorb effector molecules. Finally, when the severity of the pathogen attack increases, either by virtue of higher concentration of effectors, or by more efficient binding of effector to its target, the decoy also tends to function as a trigger rather than a sink, leaving fewer chances for the infection to pass undetected.  Our model also suggests that in principle the decoy roles of sink and trigger could  continuously adjust to the changing environment, both external and intrinsically cellular,  through  evolution of binding affinities and expression levels.  

Presently, we are unaware of the existence of any other quantitative description of the functioning of decoys in the guard-guardee model of plant immunity, whether as a sink or as a trigger. We hope that the approach developed here, which links the underlying biochemistry to resulting cost of infection and hence to evolutionary considerations, will prove to be useful both for empirical and for further theoretical studies of decoy signaling mechanisms in general, and for plant immunity in particular. For example, it would be nice to have comparative data linking parameters such as cost of maintaining decoys and binding affinities in signaling pathways to the functioning of decoys as sinks or triggers.  

Our model is based on the assumption that the host species is genetically predisposed for the production of decoys, i.e., that the mutation enabling production of decoy proteins has already occurred (e.g. through gene duplication). Under these assumptions, the model studies whether decoys should be used as sinks or as triggers, and the only scenario in which production of decoy proteins is not favoured at all occurs when such production is too costly. Thus, assuming that genetic constraints (e.g. low rate of gene duplication) do not prevent the occurrence of suitable mutations, decoy mechanisms should be expected to occur in many natural systems. The five decoy  examples presented in \cite{hoorm_kamoun_09} indicate that this mechanism may have evolved independently in phylogentically diverse plant species, and hence that decoys may indeed be a wide spread phenomenon. 
The simplicity and potential efficiency of the decoy mechanism suggests that its occurrence may not be limited to plants and that it may be used in animal immunity as well, and it remains to be seen how ubiquitous and general the use of decoys in immune response systems is.

\bibliography{decoy}
\bibliographystyle{prslb}
\end{document}